# The 2020 Global Stock Market Crash: Endogenous or Exogenous?

Ruiqiang Song[1], Min Shu[2, *], Wei Zhu[3]

[1] Michigan Technology University, Houghton, MI, USA
[2] Mathematics, Statistics & Computer Science Department, University of Wisconsin-Stout, Menomonie, WI, USA
[3] Department of Applied Mathematics & Statistics, Stony Brook University, Stony Brook, NY, USA

## Abstract

Starting on February 20, 2020, the global stock markets began to suffer the worst decline since the Great Recession in 2008, and the COVID-19 has been widely blamed on the stock market crashes. In this study, we applied the log-periodic power law singularity (LPPLS) methodology based on multilevel time series to unravel the underlying mechanisms of the 2020 global stock market crash by analyzing the trajectories of 10 major stock market indexes from both developed and emergent stock markets, including the S&P 500, the DJIA, and the NASDAQ from the United State, the FTSE from the United Kingdom, the DAX from Germany, the NIKKEI from Japan, the CSI 300 from China, the HSI from Hong Kong, the BSESN from India, and the BOVESPA from Brazil. In order to effectively distinguish between endogenous crash and exogenous crash in stock market, we proposed using the LPPLS confidence indicator as a classification proxy. The results show that the apparent LPPLS bubble patterns of the super-exponential increase, corrected by the accelerating logarithm-periodic oscillations, have indeed presented in the price trajectories of the seven indexes: S&P 500, DJIA, NASDAQ, DAX, CSI 300, BSESN, and BOVESPA, indicating that the large positive bubbles have formed endogenously prior to the 2020 stock market crash, and the subsequent crashes for the seven indexes are endogenous, stemming from the increasingly systemic instability of the stock markets, while the well-known external shocks, such as the COVID-19 pandemic, the corporate debt bubble, and the 2020 Russia–Saudi Arabia oil price war, only acted as sparks during the 2020 global stock market crash. In contrast, the obvious signatures of the LPPLS model have not been observed in the price trajectories of the three remaining indexes: FTSE, NIKKEI, and HSI, signifying that the crashes in these three indexes are exogenous, stemming from external shocks. We also found that in terms of the regime changes of the stock markets, no obvious LPPLS negative bubble pattern has been observed in the price trajectories of the 10 stock market indexes, indicating that the regime changes from a bear market to a bull market in late March 2020 are exogenous, stemming from external factors. The unprecedented market and economy rescue efforts from federal reserves and central banks across the world in unison may have played a critical role in quelling the 2020 global stock market crash in the nick of time. The novel classification method of crash types proposed in this study, endogenous or exogenous via LPPLS confidence indicator analyses, can also be used to analyze regime changes of any price trajectories in global financial markets.

Keywords: 2020 stock market crash, COVID-19, Log-periodic power law singularity (LPPLS), LPPLS Confidence indicator, Endogenous and exogenous bubble, Financial bubble and Crash.

*Correspondence to: Mathematics, Statistics & Computer Science Department, 327 Jarvis Hall-Science Wing, University of Wisconsin-Stout, Menomonie, WI, USA.
*E-mail address:* rsong1@mtu.edu (R. Song), shum@uwstout.edu (M. Shu), wei.zhu@stonybrook.edu (W. Zhu)

# 1. Introduction

Beginning on February 20, 2020, the global stock markets have turned the regime in unison from a bull market to a bear market, as shown in Figure 1. In the ensuing five weeks, the three major U.S. stock market indexes: the S&P 500, the NASDAQ composite, and the Dow Jones Industrial Average index, have plunged dramatically, falling by 33.9%, 30.1% and 37.1%, respectively. It is the worst decline since the Great Recession in 2008, interrupting the bull market trend of the past 11 years from March 2009 to February 2020. During the long booming period in the U.S. stock market since the 2008 financial crisis, the S&P 500 index has soared tremendously from 676.53 on March 9, 2009 to 3,386.15 on February 19, 2020, a four-fold increase. In the 2020 stock market crash, the level-1 trading curbs or circuit breakers due to massive panic sell-offs have been triggered repeatedly when the S&P 500 Index dropped in the range of 7% to 13%, resulting in 15-minute trading halts on major U.S. stock markets on 3/9/2020, 3/12/2020, 3/16/2020, and 3/18/2020, respectively. This is an extremely rare situation that the trading curb was triggered four times within 10 days. Since the trading curb rule was instigated after the Black Monday in 1987, the first and only other trading curb occurred on 27 October 1997, more than 22 years ago.

The 2020 Stock market Crash from February to April 2020, also called the Coronavirus Crash, has greatly impacted the lives and livelihoods of a multitude of people across the globe. During the crisis, the yield on the US 10-year Treasury note -- the most popular debt instrument and lowest-risk investment in the world backed by the full faith and credit of the United States and serving as broader barometer for economic growth, inflation expectations and financial conditions -- dropped by 65.4% within three weeks, falling from 1.56% on 2/19/2020 to 0.54% on 3/9/2020, the lowest close in history (FRED, 2020). Indeed, the Treasury yield curve rates started to inverse on 2/18/2020 with the yield on the 10-year note dropped to 1.55% while the yield on the three-month bills rose to 1.58%. As the situation grew worse, the inversion deteriorated steadily and the yield rate gap between the three-month bills and the 10-year note expanded to 15 basis points on 2/24/2020. The abnormally inverted Treasury yield curve indicated that people had lost confidence in the economic stability of the coming months or years, signaling an impending recession. To ensure banks had enough liquidity, the Federal Reserve revived the Quantitative Easing (QE) program on March 15, 2020 and announced to purchase $500 billion in U.S. Treasuries and $200 billion in mortgage-backed securities. Just one week later, the Federal Open Market Committee (FOMC) convened an emergency meeting to expanded the QE purchases to an unlimited amount (FED, 2020).

Although the proximal origins of typical market crashes in financial crises are so diverse, the root cause of financial crashes can usually be divided into two types: endogenous and exogenous. In endogenous crash, large declines in price trajectories are caused by factors within the financial market, such as the Wall Street Crash of October 1929 and the dot.com crash of April 2000, while in exogenous crash, large declines in price trajectories are induced by external shocks, such as the Nazi invasion of the western Europe on May 10, 1940, and the resignation and the subsequent controversial pardoning of President Richard Nixon due to the Watergate scandal on Augusts 8 and September 8, 1974 (Johansen & Sornette, 2010).

Thus far, multiple external shocks have been widely blamed on the 2020 global stock market crash. The outbreak of the novel coronavirus COVID-19 in January 2020 had attracted global attention. On January 30, 2020, the Director-General of the World Health Organization (WHO) announced that the outbreak of the COVID-19 constituted a Public Health Emergency of International Concern (PHEIC), the highest level of alarm of the WHO. Deeply concerned about the shocking speed of disease spread, severity, and government inaction, the WHO declared the disease a global pandemic on March 11, 2020 (WHO, 2020). Since the 1918 flu pandemic, the COVID-19 pandemic has been the most impactful pandemic.. Albuquerque et al. (2020) argued that the COVID-19 pandemic and the subsequent lockdowns caused an external and unparalleled shock that resulted in the 2020 global stock market crash. Hence the 2020 Global Stock Market Crash is also called the Great Coronavirus Crash (Coy, 2020). Besides the COVID-19 pandemic-induced market instability, the mammoth nonfinancial corporate debt bulge had showed signs of financial market stress and liquidity crisis, which was also found responsible for the economic recession (Lynch, 2020). Further, Guardian (2020) believed that the 2020 Russia–Saudi Arabia oil price war together with the Coronavirus outbreak had triggered the global stock market crash.

In this study, we used the Log Periodic Power Law Singularity (LPPLS) model (Drozdz et al., 1999; Feigenbaum & Freund, 1996; Johansen et al., 2000; Johansen et al., 1999; Sornette & Johansen, 2001; Sornette et al., 1996) to disclose the underlying mechanisms of the 2020 Stock market Crash. The LPPLS model originates from the interface of financial economics, behavioral finance and statistical physics by combining the economic theory of rational expectations, behavioral finance of herding of traders, and the mathematical and statistical physics of bifurcations and phase transitions. In the LPPLS model, a positive (negative) financial bubbles is represented as a process of unsustainably super-exponential growth (decline) to achieve an infinite return in finite time, forcing a short-lived correction molded according to the symmetry of discrete scale invariance (Sornette, 1998). The correction of the price in a finite time singularity can be a boom (or crash) as the price increase (or drop) significantly, or a change of regime as the price change velocity increase (or decrease) dramatically in a short time scale. The LPPLS model can capture two distinct characteristics normally observed in the regime of bubbles, that is, faster-than-exponential growth of the price resulting from positive feedback by imitation and herding behavior of noise traders, and the accelerating log-periodic volatility fluctuations of the price growth from expectations of higher returns or an upcoming crash. We found that these two significant features can be clearly observed in the 10 stock market indexes shown in Figure 1, in which the transient super-exponential growth rates can be visually diagnosed as an upward curvature in the linear-log plot, adorned with significant log-periodic undulations.

In the recent years, the LPPLS model has been studied by many scholars to diagnose speculative bubbles through analyzing the LPPLS signatures embedded in the asset price trajectories. Filimonov and Sornette (2013) converted the LPPLS formulation to reduce the calibration complexity by decreasing the number of nonlinear parameters from four to three. Sornette et al. (2015) examined the real-time prediction of bubble crash in the 2015 Shanghai stock market using the LPPLS Confidence indicator and Trust indicator. Filimonov et al. (2017) adopted the adjusted profile likelihood inference method to calibrate the LPPLS model and to infer the critical time. Demos and Sornette (2017) discovered that the beginning time of a bubble is much better constrained than its end time. Demirer et al. (2019) probed the predictive power of market-based indicators by using the LPPLS confidence multi-scale indicators and found that short selling and liquidity both contributed to the bubble indicators. Shu and Zhu (2020b) proposed an adaptive multilevel time series detection methodology to enable the real-time detections of financial bubbles and crashes. In addition, the LPPLS model has been deployed to detect bubbles and crashes in a variety of financial markets including the real estate bubbles (Zhou & Sornette, 2003, 2006, 2008), the oil market bubbles (Sornette et al., 2009), the Chinese stock market bubbles (Jiang et al., 2010; Li, 2017; Shu, 2019; Shu & Zhu, 2019, 2020a; Sornette et al., 2015), and the Bitcoin bubbles (Gerlach et al., 2019; Shu & Zhu, 2020b).

It should be noted that the LPPLS model could only detect the endogenous crashes but not the exogenous crashes. This is because the LPPLS model is rooted on the faster-than-exponential growth of the price trajectory, which is a significant feature when a financial bubble results from the self-reinforcements of cooperative herding and imitative behaviors through interactions of market players involving long-memory processes of an endogenous organization. In an exogenous crash, the exogenous shocks can change an asset's fundamental value underlying a financial market and leave some observable signatures in the price trajectories; however, the signature of the LPPLS model before a crash will be absent from the corresponding price trajectories.

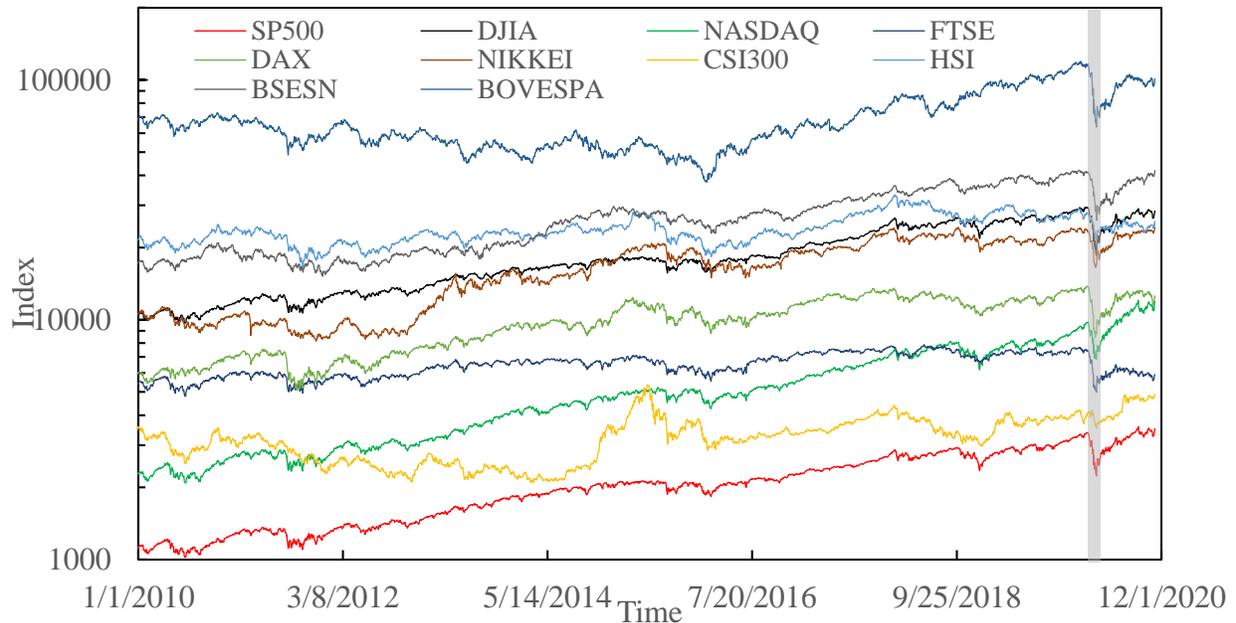

Figure 1. Evolution of price trajectories of the 10 major stock market indexes from January 2010 to November 2020 (The dark shadow box indicates the period of 2020 global stock market crash)

We adopted the LPPLS model in this study to investigate the 2020 global stock market crash by analyzing the performance of the 10 major global stock market indexes from both developed and emergent stock markets, including the S&P 500 (SP500), the Dow Jones Industrial Average (DJIA), and the Nasdaq Composite (NASDAQ) indexes for the stock market exchanges in United State, the Financial Times Stock Exchange 100 Index (FTSE) for the London Stock Exchange in United Kingdom, the DAX performance index (DAX) for the Frankfurt Stock

Exchange in Germany, the Nikkei Stock Average index (NIKKEI) for the Tokyo Stock Exchange in Japan, the CSI 300 stock market index (CSI300) for the Shanghai Stock Exchange and the Shenzhen Stock Exchange in China, the Hang Seng Index (HSI) for the Hong Kong Stock Exchange in Hong Kong, the BSE SENSEX index (BSESN) for the Bombay Stock Exchange in India, and the Bovespa Index (BOVESPA) for the B3 Stock Exchange in Brazil. Figure 1 shows the evolution of price trajectories of these 10 major stock market indexes from January 2010 to November 2020. We also proposed a novel classification method based on the LPPLS confidence indicator to analyze crash types and regime changes of price trajectories in stock market. In order to investigate the 2020 global stock market crash from different time scales, the multilevel time series of price trajectories were employed in this study. The paper is organized as the following: Section 2 presents our methodology; Section 3 discusses the analyses and the results; and, finally, Section 4 concludes.

## 2. Methodology

### 2.1 The Log-Periodic Power Law Singularity (LPPLS) Model

Asset price trajectories in bubble regimes usually show two typical characteristics: the transient faster-than-exponential growth decouples from the intrinsic fundamental value, and the accelerating log-periodic volatility fluctuates near the crash points. Based on the rational expectation bubble model (Blanchard & Watson, 1982), the LPPLS model, also known as Johansen-Leoit-Sornette model, captures these two typical characteristics to model a financial bubble as a process of faster-than-exponential power law growth intervened by short-lived corrections organized according to the symmetry of discrete scale invariance (Sornette, 1998). The LPPLS model integrates the rational expectation economic theory, the imitation and herding behavioral finance by investors and traders, and the mathematical and statistical physics of bifurcation and phase transitions (Yan et al., 2010). In the LPPLS model, the dynamics of the observed asset price $p(t)$ is described as:

$$\frac{dp}{p} = \mu(t)dt + \sigma(t)dW - kdj \tag{1}$$

where $\mu(t)$ is the time-dependent expected return, $\sigma(t)$ is the volatility, $dW$ is the infinitesimal increment of a Wiener process with zero mean and unit variance, $dj$ represents a discontinuous jump with value before the crash set to 0 and after the crash set to 1, while the term $k$ quantifies the loss amplitude of a possible crash. The dynamics of the jumps are determined by a crash hazard rate $h(t)$, which measures the crash probability at a specified time $t$. Under the condition that no jump has occurred, the expected value of $dj$ between $t$ and $t + dt$ can be calculated as $E[dj] = h(t)dt$. The LPPLS model considers two types of agents: (a) the rational traders with rational expectations, and, (b) the noise traders susceptible to exhibit imitation and herding behavior. The LPPLS model assumes that the collective behavior of noise traders can destabilize asset prices through related transactions. The aggregate effect of noise traders can be determined by the following dynamics of the crash hazard rate $h(t)$ (Johansen et al., 2000):

$$h(t) = \alpha(t_c - t)^{m-1}(1 + \beta cos(\omega \ln(t_c - t) - \phi')) \tag{2}$$

where $\alpha, \beta, \omega$, and $t_c$ are model parameters. From Equation (2), the crash risk at a specified time $t$ resulting from the imitation and herding behavior is a sum of the power law singularity $\alpha(t_c - t)^{m-1}$ which embodies the positive feedback mechanism of the herding behaviors of the noise traders, modulated by large-scale amplitude oscillations that are periodic in the logarithm of the time to the critical time $t_c$. The term $\cos(\omega \ln(t_c - t) - \phi')$ signifying the existence of a possible hierarchical cascade of panic acceleration punctuating the bubble growth, results either from a preexisting hierarchy in noise trader sizes (Sornette & Johansen, 1997) and/or from the interplay between market price impact inertia and nonlinear fundamental value investing (Ide & Sornette, 2002).

Given the non-arbitrage condition, the unconditional expectation $E[dp]$ of the price increment should be 0, resulting in:

$$\mu(t) \equiv E[\frac{dp/dt}{p}]_{\text{no crash}} = kh(t) \tag{3}$$

Under the condition that no crash has yet occurred, the simple mathematical formulation of the LPPLS for the expected trajectory of the log-price in a bubble regime can be achieved by solving Equation (1) while incorporating Equations (2) and (3) (Sornette, 2003):

$$\text{LPPLS}(t) \equiv \ln E[p(t)] = A + B(t_c - t)^m + C(t_c - t)^m \cos[\omega \ln(t_c - t) - \phi] \tag{4}$$

where $B = -k\alpha/m$ and $C = -k\alpha\beta/\sqrt{m^2 + \omega^2}$. By expanding the term $C \cos[.]$ with two linear parameters $C_1 = C\cos\phi$ and $C_2 = C\sin\phi$ to replace the linear parameter $C$ and the nonlinear parameter $\phi$, the mathematical formulation of the LPPLS in Equation (4) can be rephrased as (Filimonov & Sornette, 2013):

$$\begin{aligned}\text{LPPLS}(t) \equiv E_t[\ln p(t)] = A &+ B(t_c - t)^m + C_1(t_c - t)^m \cos[\omega \ln(t_c - t)] \\ &+ C_2(t_c - t)^m \sin[\omega \ln(t_c - t)]\end{aligned} \tag{5}$$

Here $A > 0$ is the expected value of the log-price at $t_c$. The bubble regime is usually characterized by $0 < m < 1$ and $B < 0$ ($B > 0$) to ensure that the price changes super-exponentially as time goes towards $t_c$. The condition of $m > 0$ indicates that the price remains finite at the $t_c$, while $m < 1$ ensures that a singularity exists. It should be noted that the critical time $t_c$ is the most probable time for a regime change based on the asset price trajectories, which may be either a large spike or crash, or a change of the growth rate from super-exponential to exponential or lower growth, and the subsequent termination of the accelerated oscillations.

## 2.2 Calibration of the Model

The remodeled LPPLS formula in Equation (5) has 3 nonlinear parameters $\{t_c, m, \omega\}$ and 4 linear parameters $\{A, B, C_1, C_2\}$. Using the $L^2$ norm, the sum of squares of residuals of the converted LPPLS formula can be written as:

$$F(t_c, m, \omega, A, B, C_1, C_2) = \sum_{i=1}^{N} [\ln p(\tau_i) - A - B(t_c - \tau_i)^m - C_1(t_c - \tau_i)^m \cos(\omega \ln(t_c - \tau_i)) \\ - C_2(t_c - \tau_i)^m \sin(\omega \ln(t_c - \tau_i))]^2 \tag{6}$$

where $\tau_1 = t_1$ and $\tau_N = t_2$. The cost function $\chi^2(t_c, m, \omega)$ can be obtained by relating the 4 linear parameters $(A, B, C_1, C_2)$ to the 3 nonlinear parameters $(t_c, m, \omega)$:

$$\chi^2(t_c, m, \omega) = F_1(t_c, m, \omega) = \min_{\{A,B,C_1,C_2\}} F(t_c, m, \omega, A, B, C_1, C_2) = F(t_c, m, \omega, \hat{A}, \hat{B}, \hat{C}_1, \hat{C}_2) \quad (7)$$

where the hat symbol ^ represents the estimated parameters. The 4 linear parameters $(A, B, C_1, C_2)$ can be estimated by solving the optimization equation:

$$(\hat{A}, \hat{B}, \hat{C}_1, \hat{C}_2) = arg \min_{(A,B,C_1,C_2)} F(t_c, m, \omega, A, B, C_1, C_2) \quad (8)$$

Equation (8) can be solved analytically through the following matrix equations

$$\begin{pmatrix} N & \sum f_i & \sum g_i & \sum h_i \\ \sum f_i & \sum f_i^2 & \sum f_i g_i & \sum f_i h_i \\ \sum g_i & \sum f_i g_i & \sum g_i^2 & \sum h_i g_i \\ \sum h_i & \sum f_i h_i & \sum g_i h_i & \sum h_i^2 \end{pmatrix} \begin{pmatrix} \hat{A} \\ \hat{B} \\ \hat{C}_1 \\ \hat{C}_2 \end{pmatrix} = \begin{pmatrix} \sum \ln p_i \\ \sum f_i \ln p_i \\ \sum g_i \ln p_i \\ \sum h_i \ln p_i \end{pmatrix} \quad (9)$$

where $f_i = (t_c - t_i)^m$, $g_i = (t_c - t_i)^m \cos(\omega \ln(t_c - t_i))$, and $h_i = (t_c - t_i)^m \sin(\omega \ln(t_c - t_i))$.

The three nonlinear parameters $\{t_c, m, \omega\}$ can be estimated by solving the resulting nonlinear optimization problem:

$$\{\hat{t}_c, \hat{m}, \hat{\omega}\} = arg \min_{\{t_c, m, \omega\}} F_1(t_c, m, \omega) \quad (10)$$

The seven parameters $(t_c, m, \omega, A, B, C_1, C_2)$ in the transmuted LPPLS model can be estimated by calibrating the model based on the price trajectory using the Ordinary Least Squares method. The covariance matrix adaptation evolution strategy (CMA-ES) developed by Hansen et al. (1995) is used to solve this optimization problem. The best estimation of the three nonlinear parameters $(t_c, m, \omega)$ can be obtained by minimizing the sum of residuals between the fitted LPPLS model and the observed price trajectory.

## 2.3 LPPLS Confidence Indicator

In order to measure the sensitivity of the detected bubble pattern to the selection of the start time $t_1$ in the fitting windows, Sornette et al. (2015) proposed the LPPLS confidence indicator which is defined as the fraction of fitting windows where the calibrated LPPLS models meet the specified filter conditions. Larger LPPLS confidence indicator simply means more fitting windows having the signatures of the LPPLS model and the hence the detected LPPLS bubble patterns are more reliable. A small value of the LPPLS confidence indicator indicates a possible fragility because only few fitting windows would show the LPPLS bubble pattern.

For a specified data point $t_2$ corresponding to a fictitious "present", the LPPLS confidence indicator can be determined in the following way: (a) create the total fitted time windows by moving $t_1$ toward the $t_2$ with a step of $dt$, (b) calibrate the LPPLS model in the specified search space for each fitting time window, and (c) divide the number of time windows satisfying the specified filter conditions by the total number of the fitting windows. It should be noted that the LPPLS confidence indicator is causal because it is calculated based on the data prior to $t_2$.

Given a specified endpoint $t_2$, a group of time series of asset price trajectories in this study is generated by shrinking the length of the time window $(t_1, t_2)$ from 650 data points to 30 data points in the step of 5 data points, thereby creating 125 fitting windows for each $t_2$. To ensure model rigorousness, the search space is set to (Shu & Zhu, 2020b):

$$m \in [0,1], \omega \in [1, 50], t_c \in \left[t_2, t_2 + \frac{t_2-t_1}{3}\right], \frac{m|B|}{\omega\sqrt{C_1^2+C_2^2}} \geq 1 \tag{11}$$

The condition $t_c \in [t_2, t_2 + (t_2 - t_1)/3]$ ensures that the critical time $t_c$ can be predicted not only after the endpoint $t_2$, but also adjacent to $t_2$ because the predictability of the LPPLS model is degraded while moved from $t_2$ (Jiang et al., 2010). Here the Damping parameter $m|B|/\left(\omega\sqrt{C_1^2+C_2^2}\right) \geq 1$ indicates that the crash hazard rate $h(t)$ is non-negative (Bothmer & Meister, 2003).

To determine the valid LPPLS fits, the calibrated LPPLS model will be filtered under the following conditions (Shu & Zhu, 2020b):

$$m \in [0.01, 0.99], \omega \in [2, 25], t_c \in \left[t_2, t_2 + \frac{t_2 - t_1}{5}\right], \frac{\omega}{2} \ln\left(\frac{t_c - t_1}{t_c - t_2}\right) \geq 2.5,$$
$$\max\left(\frac{|\hat{p}_t - p_t|}{p_t}\right) \leq 0.20, \ p_{lomb} \leq \alpha_{sign}, \ln(\hat{p}_t) - \ln(p_t) \sim AR(1) \tag{12}$$

The filter conditions are derived from empirical evidence collected from previous investigations on financial bubbles (Jiang et al., 2010; Sornette et al., 2015). The condition of $(\omega/\pi)\ln[(t_c - t_1)/(t_c - t_2)] \geq 2.5$ is used to distinguish between genuine log-periodic signals and noise-generated signals (Huang et al., 2000). The condition of $\max(|\hat{p}_t - p_t|/p_t) \leq 0.20$ ensures that the fitted price $\hat{p}_t$ is close to the actual asset price $p_t$. The condition $P_{lomb} \leq \alpha_{sig}$ assures the existence of the logarithm-periodic oscillations in the fitted LPPLS model by applying the Lomb spectral analysis to the time series of the detrended residual $r(t) = (t_c - t)^{-m}(\ln[p(t)] - A - B(t_c - t)^m)$ (Sornette & Zhou, 2002). The $\ln(\hat{p}_t) - \ln(p_t) \sim AR(1)$ condition implies that the LPPLS fitting residuals can be modeled by a mean-reversal Ornstein-Uhlenbeck (O-U) process when the logarithmic price in the bubble regime is due to a deterministic LPPLS component (Lin et al., 2014). Only calibration results meeting the filter conditions in Equation (12) are believed to be valid while the rest are deemed superfluous.

### 2.4 Classification of Crash Types

Because the LPPLS model, modeling the transient super-exponential growth of asset trajectories resulting from self-reinforcing cooperative herding and imitative behaviors through interactions between market participants involving long-term memory processes of an endogenous organization, can only detect the endogenous crashes, we propose using the confidence indicator as a classification proxy to distinguish between endogenous crash and exogenous crash in the financial markets. In order to ensure the classification robustness of crash types and reduce the classification errors due to model over-fitting and other potential issues in the calibration process, a series of specified LPPLS confidence indicator values based on the LPPLS fitting windows and the time series attributes of the asset price trajectory can be used as thresholds to identify the crash types. As a rule of thumb, a value LPPLS confidence interval greater than 5% for the daily price trajectory or 2% for weekly price trajectory, signals that the price process is unsustainable and bears a substantial risk for an impending critical transition, thereby the value of 5% and 2% are adopted in this study as the threshold for the daily and weekly price trajectory, respectively.

Given a specified time point, if the associated confidence indicator value exceeds the specified threshold value, we conclude that the signatures of the LPPLS model have been identified in the corresponding price trajectory, and the price trajectory is indeed in an endogenous bubble state reaching a state of system instability, and thus any small disturbances would have the potential to trigger a mature bubble to burst resulting in an endogenous crash. In contrast, if the confidence indicator value at a given time point is lower than the specified threshold, the evidence is not enough to conclude that the price trajectory clearly contain the signature of LPPLS model, and so the price trajectory may not be in an endogenous bubble state. Small disturbances will not be able to trigger the crash of a price trajectory outside a bubble status. The crash of a price trajectory in a state of system stability can only be caused by large external shocks resulting in an exogenous crash.

## 3. Empirical Analysis

### 3.1 Bubble Identification Based on Daily Data

In this section, we have collected the daily data of the 10 major global stock market indexes from both developed and emergent stock markets, including S&P 500, DJIA, NASDAQ, FTSE, DAX, NIKKEI, CSI 300, HSI, BSESN, and BOVESPA. The stock market data used in this study came from Yahoo Finance (https://finance.yahoo.com/). To calculate the LPPLS confidence indicator, we shrunk the length of time windows $t_2 - t_1$ from 650 trading days to 30 trading days in steps of 5 trading days, and moved the endpoint $t_2$ from January 2, 2019 to November 6, 2020. The LPPLS confidence indicator for a specified time is causal because it is computed using the prior data only. Both the positive and the negative bubbles in this study were detected based on the daily data of the stock market indexes. The positive bubble is related to accelerating price increase and is vulnerable to regime changes in the form of crashes or volatile sideway plateaus. In contrast, the negative bubble is associated with accelerating price decrease, and is prone to regime changes in the form of rallies or volatile sideway plateaus.

Figure 2 shows the LPPLS confidence indicator for positive bubbles in red and negative bubble in green along with the index price in blue for the 10 major stock market indexes based on daily data from January 2, 2019 to November 6, 2020. We can intuitively perceive the confidence level of the observed LPPLS bubble patterns in Figure 2, because the sensitivity of the bubble pattern to the start time point is measured by the LPPLS confidence indicator. The confidence indicator can reach a high enough value, say 10%, when the LPPLS bubble patterns exist in a substantial number of fitting windows for a specified pseudo-present time, indicating the detected LPPLS bubble pattern is reliable and not sensitive to the choice of the start time points -- thereby the price trajectory in bubble regime can be confirmed. In contrast, the confidence indicator can be low close to zero, say 1%, when the bubble patterns are only observed in few fitting time windows for a specified pseudo-present time, which indicates the detected LPPLS bubble pattern is sensitive to the choice of the start time points and the over-fitting risk may exist -- therefore the price trajectory is less likely to be in a bubble regime.

Figure 2 (a) shows the detected SP500 bubble status, including three obvious clusters of positive bubbles: between December 12, 2019 and January 29, 2020, between February 10 and February 21, 2020, and between August 6 and September 4, 2020, and one subtle cluster of negative bubble between March 17 and March 27, 2020. The positive confidence indicator reaches the maximum value of 16% on Feb 19, 2020 in the cluster between February 10 and February 21, 2020, which means that the 20 out of 125 fitting windows can successfully pass the filter conditions. It indicates that the detected LPPLS pattern is reliable, and the positive bubble status on February 19, 2020, can be confirmed. It is highly possible that the accelerating increase trend of the SP500 index is not sustainable, and the regime change of the SP500 index is prone to occur. The prediction was confirmed by the fact that the SP500 dropped dramatically from 3,386.1 on February 19, 2020 to 2,237.4 on March 23, 2020, losing 33.9% of its value within 24 trading days. In the positive bubble cluster between August 6 and September 4, 2020, the positive confidence indicator reaches the peak of 8.8% on September 3, 2020, corresponding to the fact that the SP500 decreased by 9.6% from 3,580.8 on September 2, 2020 to 3,236.9 on September 23, 2020.

During the period of the 2020 stock market crash, similar detected patterns of positive bubble shown in Figure 2 (a) for the S&P 500 can also be found in other subfigures in Figure 2, including: (b) DJIA, (c) NASDAQ, (e) DAX, (g) CSI300, (i) BSESN, and (j) BOVESPA. It should be noted that a subtle cluster of positive bubble in Figure 2 (d) FTSE can be observed from December 24, 2019 to February 13, 2020 with the peak value of 0.8%. The confidence indicator value of 0.8% represents that only 1 out of 125 fitting windows can successfully pass the filter conditions, indicating that the detected LPPLS positive bubble pattern is not reliable, and the over-fitting risk may exist. The similar patterns of a hint of a positive bubble like that shown in Figure 2 (d) can also be observed in two other subfigures, namely, Figure 2 (f) NIKKEI and (h) HSI.

In Figure 2, we can notice that clusters of positive bubbles sometimes overlap with clusters of negative bubbles. For example, as shown in Figure 2 (h) HSI, a subtle cluster of positive bubbles can be observed from July 17, 2020 to September 3, 2020 with the peak confidence indicator of 1.6%, while a noticeable cluster of negative bubble exists in the same time frame with the peak confidence indicator of 6.4%. The overlap of different types of clusters is due to the fact that an asset price trajectory may be at a trough on a short-term time scale while approaching a peak on a long-term scale, and vice versa.

Table 1 summarizes the statistics of positive bubble detection results for the 10 stock market indexes based on daily data during the COVID crash from February to April 2020 and the related information of the peaks and valleys. It can be observed that the BOVESPA index in Brazil stock market has the largest crash size of 45.4%, followed by the DAX index in Germany stock market with the crash size of 38.8%. Surprisingly, 8 out of 10 indexes lost more than 30% of their values within five weeks in this crash. The CSI300 index in Chinese stock market and HSI index in Hongkong stock market suffered relatively less loss, dropping 16.1% and 21.5%, respectively.

In this study, the 5% confidence indicator value based on the empirical analysis was used as the threshold to classify crash types for short-term analysis where the bubble signals are expected to be stronger than those in the long-term analysis. The peak confidence indicator (CI) values and the types of crash for the 10 major stock market indexes during the 2020 global stock market crash are listed in Table 1. Seven out of 10 indexes, including SP500, DJIA, NASDAQ, DAX, CSI300, BSESN, and BOVESPA, have a peak confidence indicator value exceeding the 5% threshold during the COVID crash, signifying that the price trajectories of these seven indexes clearly feature the signatures of the LPPLS model endogenous bubble state, indicating that the subsequent crashes in these seven indexes during the 2020 global stock market crash are endogenous stemming from the increasingly systemic instability of the stock markets. In contrast, the peak confidence indicator values for the remaining three indexes: FTSE, NIKKEI, and HSI, are far less than the threshold value of 5%, indicating the absence of clear LPPLS bubble signatures in these price trajectories, and hence the subsequent crashes are exogenous stemming from the large external shocks, such as, the COVID-19 pandemic-induced market instability, the mass hysteria, the corporate debt bubble, etc.

Table 1: Statistics of positive bubble detection based on daily data during the 2020 global stock market crash.

| Index | Region | Peak Date | Peak Price | Valley date | Valley Price | Crash Size | Peak CI | Type of Crash* |
|---|---|---|---|---|---|---|---|---|
| SP500 | USA | 2/19/2020 | 3386.1 | 3/23/2020 | 2237.4 | 33.9% | 16.0% | Endogenous |
| DJIA | USA | 2/12/2020 | 29551.4 | 3/23/2020 | 18591.9 | 37.1% | 21.6% | Endogenous |
| NASDAQ | USA | 2/19/2020 | 9817.2 | 3/23/2020 | 6860.7 | 30.1% | 12.0% | Endogenous |
| FTSE | UK | 2/19/2020 | 7457.0 | 3/23/2020 | 4993.9 | 33.0% | 0.8% | Exogenous |
| DAX | Germany | 2/19/2020 | 13789.0 | 3/18/2020 | 8441.7 | 38.8% | 8.0% | Endogenous |
| NIKKEI | Japan | 2/12/2020 | 23861.2 | 3/19/2020 | 16552.8 | 30.6% | 0.8% | Exogenous |
| CSI300 | China | 3/5/2020 | 4206.7 | 3/23/2020 | 3530.3 | 16.1% | 8.8% | Endogenous |
| HSI | Hongkong | 2/19/2020 | 27655.8 | 3/23/2020 | 21696.1 | 21.5% | 2.4% | Exogenous |
| BSESN | India | 2/19/2020 | 41323.0 | 3/23/2020 | 25981.2 | 37.1% | 6.4% | Endogenous |
| BOVESPA | Brazil | 2/19/2020 | 116518.0 | 3/23/2020 | 63570.0 | 45.4% | 12.0% | Endogenous |

*Note: To ensure the classification robustness, the 5% confidence indicator value is used here as the threshold.

An interesting finding in Figure 2 is that most of the stock market indexes during the 2020 global stock market crash dropped to the valley at the same day on March 23, 2020. Since then, the price trajectories began to rise, and the global stock markets changed the regime from the bear market to the bull market. However, most of the confidence indicator cluster of negative bubble during this period shown in Figure 2 have peak values far less than the threshold of 5%, indicating that there are no apparent LPPLS negative bubble pattern presented in the price trajectories of these 10 stock market indexes and the subsequent regime changes are exogenous resulting from external factors. On March 23, 2020, the finance ministers and central bank executives from the world's 20 largest economies (G20) agreed to develop a joint action plan to address the economic effects of the COVID-19 pandemic and to help stabilize public confidence and the markets (Japantimes, 2020). On the same day, the Federal Reserve had committed to using its full range of tools to support the U.S. economy and announced that the FOMC would purchase Treasury securities and agency mortgage-backed securities in unlimited amounts to stabilize market functioning (FED, 2020). Under the condition that assets are difficult to sell, such purchases not only injected cash into the economy and helped restore normal functioning of key market, but also conveyed that the Fed stood ready to backstop important parts of the financial system (Ihrig et al., 2020). Other external shock events on March 23, 2020 included the Reserve Bank of India's injection of ₹1 trillion in repo operations (RBI, 2020), and the purchase of 800 billion yen of government bonds by the Bank of Japan on the open market (Mogi, 2020),. The federal reserves and central banks across the world may have played a critical role in quelling the 2020 global stock market crash, by adopting extensive measures to provide investors and markets with unprecedented supports, such as injecting cash into the economy, cutting their interest rates, and so forth.

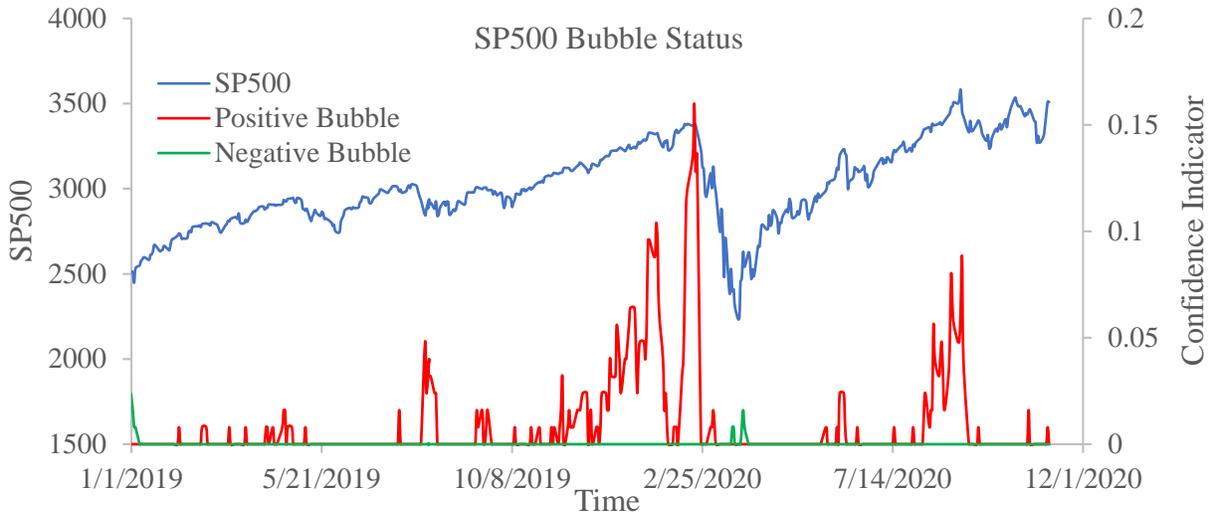

(a)

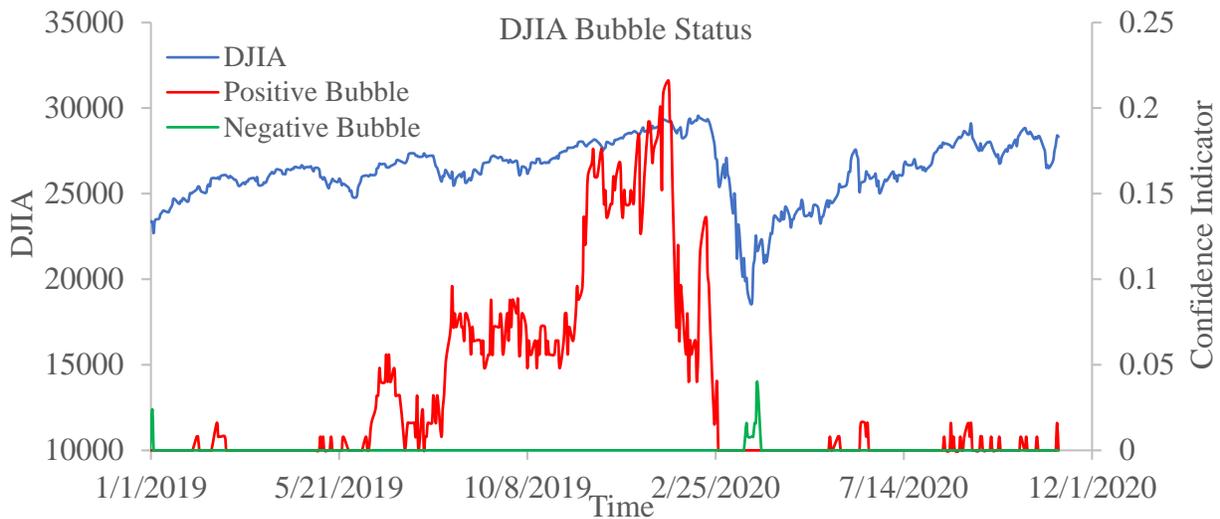

(b)

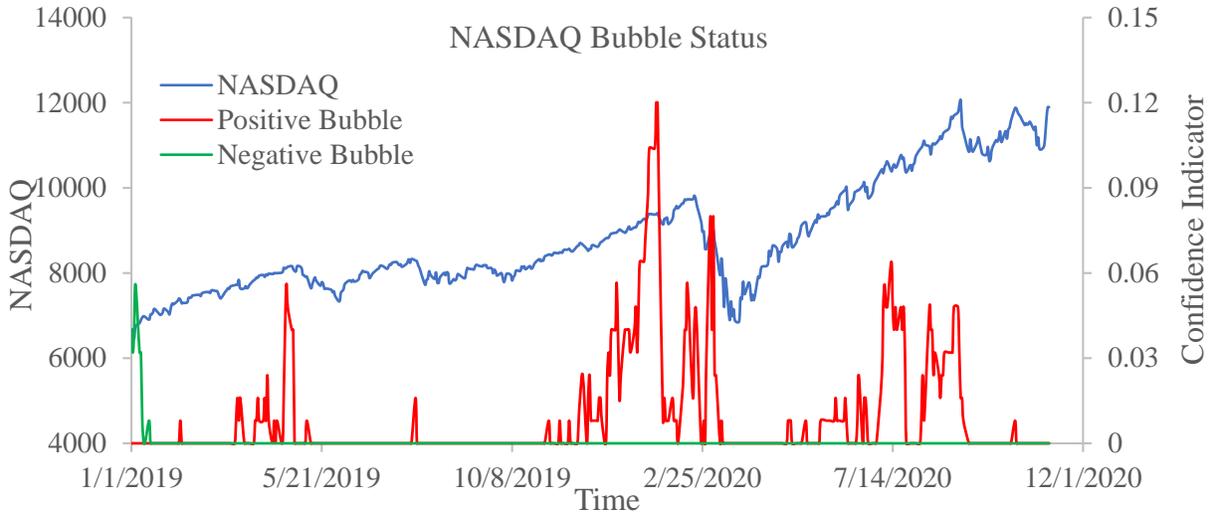

(c)

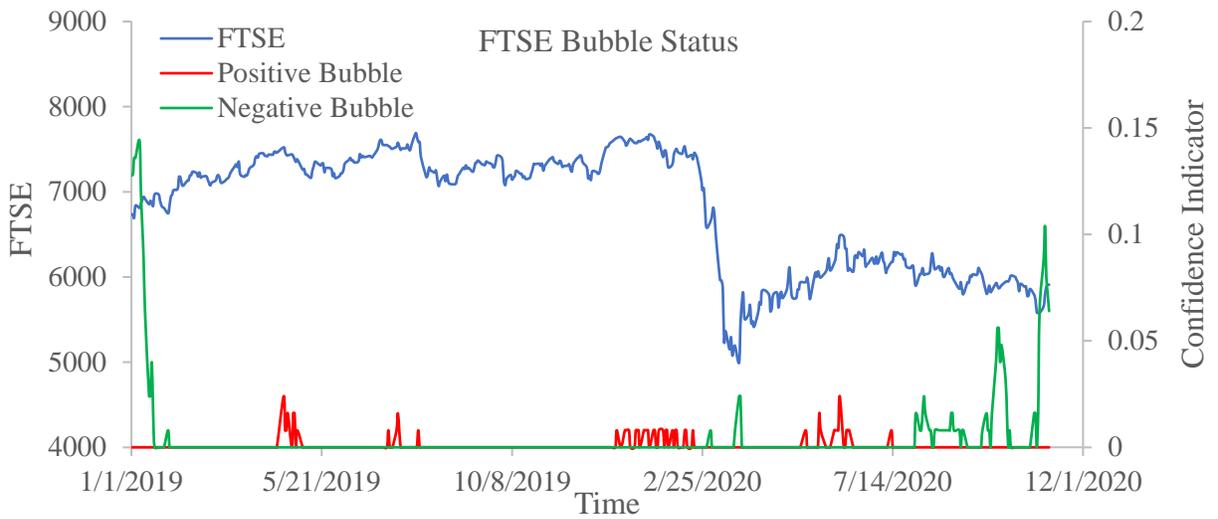

(d)

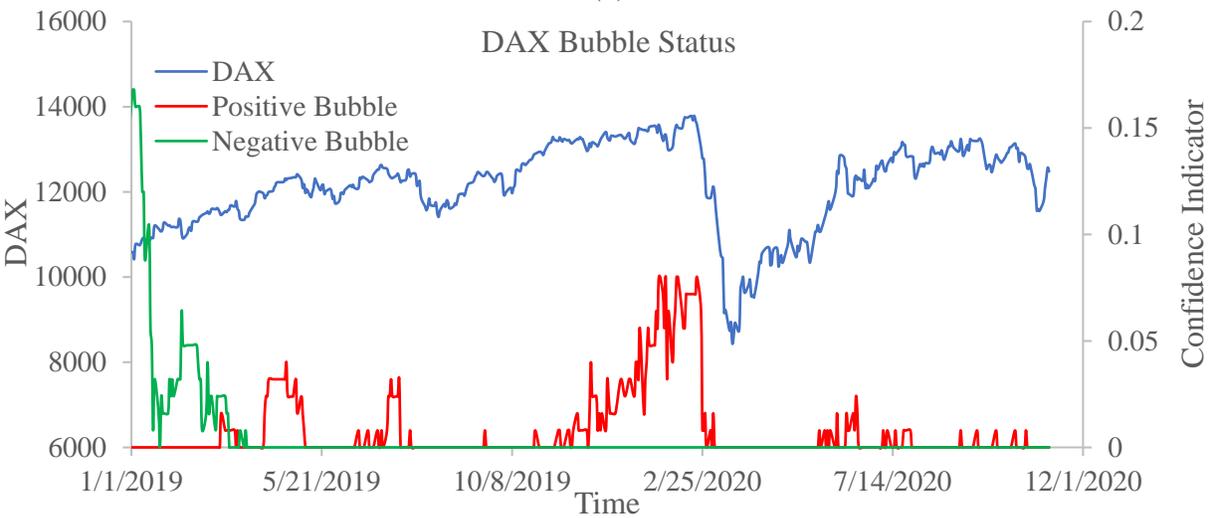

(e)

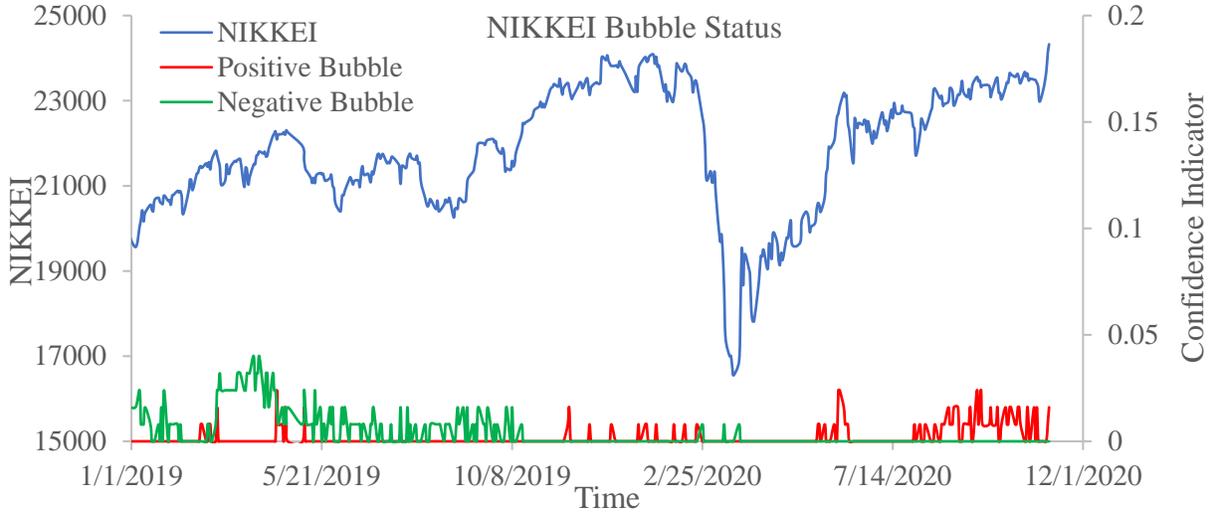

(f)

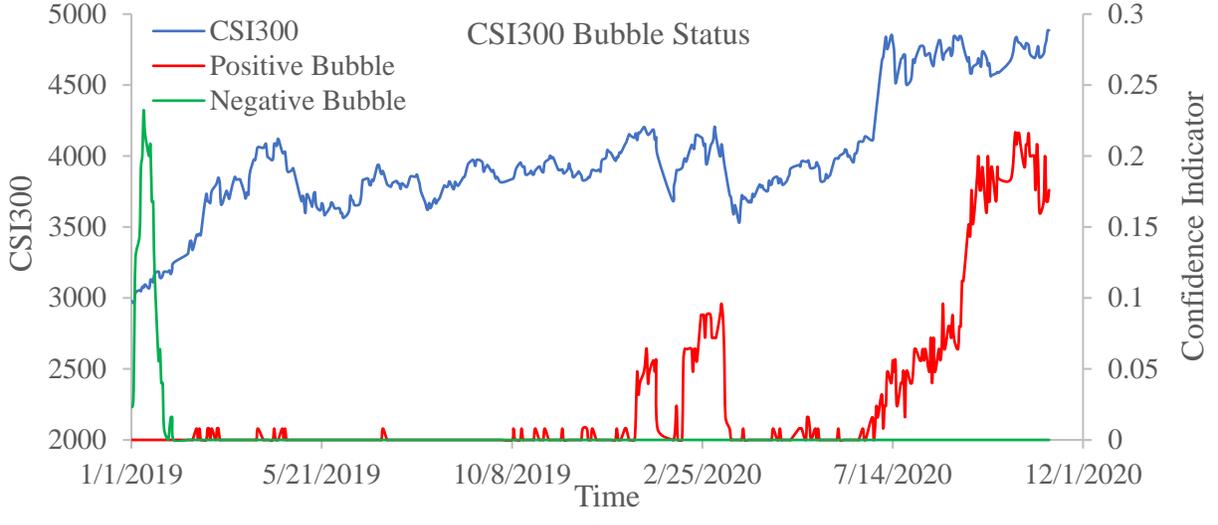

(g)

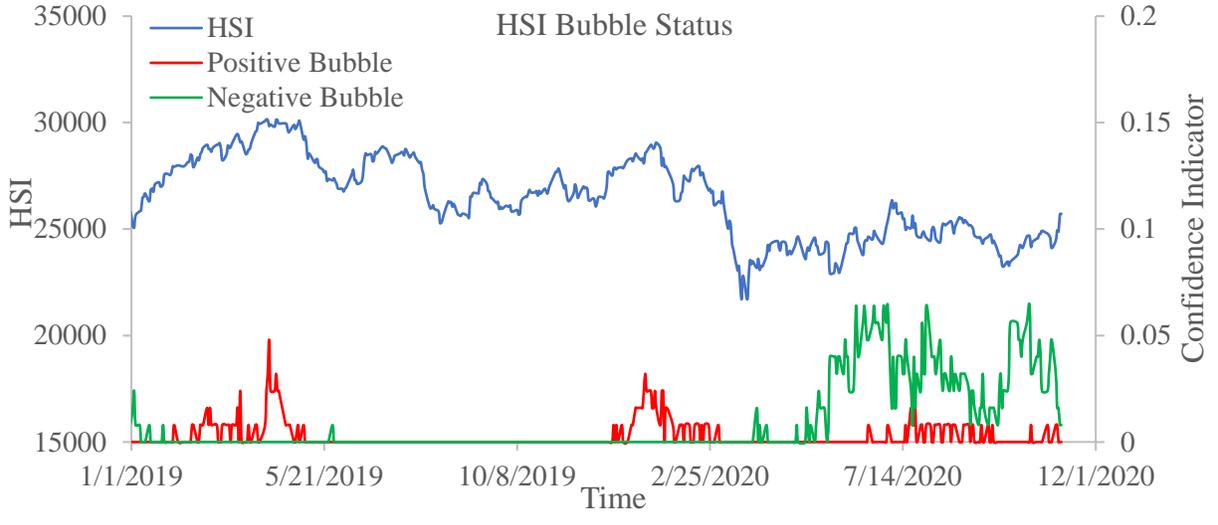

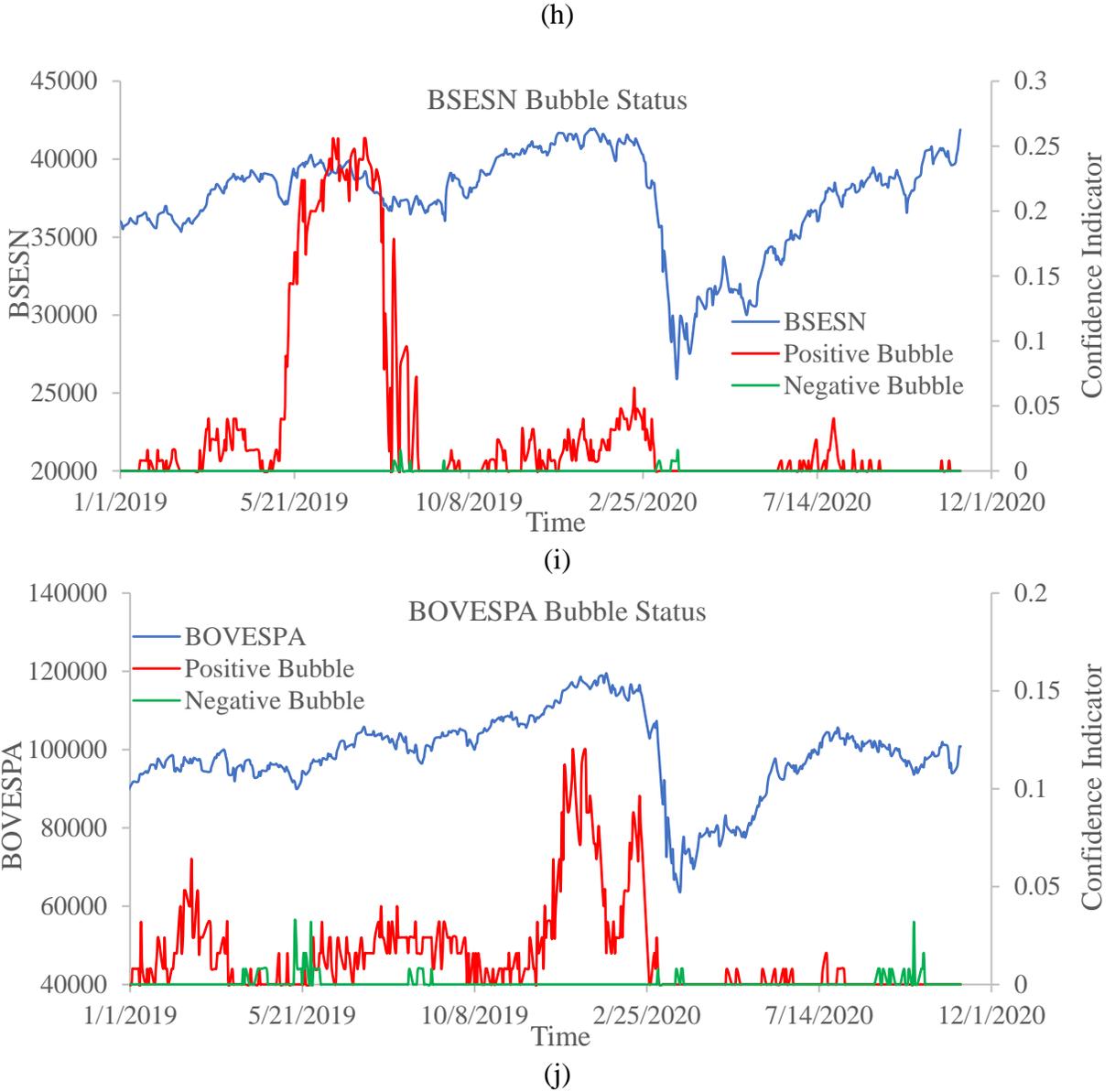

Figure 2. LPPLS confidence indicator for positive bubbles is shown in red and negative bubbles in green (right scale) along with the index price in blue (left scale), for the 10 major stock market indexes based on daily data from January 2019 to November 2020.

## 3.2 Bubble Identification Based on Weekly Data

In this section, we investigated the 2020 global stock market crash based on the weekly data from a long-term time scale. The weekly data of price trajectory was extracted in the interval of the 5 trading days from the daily trading data. Based on the extracted weekly price trajectory, we shrunk the length of time windows $t_2 - t_1$ from 650 data points to 30 data points in steps of 5 data points, and moved the endpoint $t_2$ from January 8, 2019 to November 9, 2020 to calculate the LPPLS confidence indicator. The bubble status of stock market index for a specified time point would be detected by using the data of the previous 30 to 650 data points, which corresponds to about 300 to 6500 calendar days, that is, about ten months to eighteen years.

Figure 3 shows the LPPLS confidence indicator for positive bubbles in red and negative bubble in green along with the index price in blue for the 10 major stock market indexes based on weekly data from January 8, 2019 to November 9, 2020. By using the previous data from about ten months to eighteen years before a specified time point, Figure 3 provides the useful insights into the bubble status from a long-term time scale. Figure 3 (a) presents the bubble status for the SP500 index. An obvious cluster of positive bubbles formed between January 21, 2020 and March 11, 2020 with the peak value of 16% on February 19, 2020 as shown in Figure 3 (a). This demonstrates that the S&P 500 index during the 2020 stock market crash presented a significant signature of the LPPLS model, and a positive bubble had been formed endogenously even detectable by using the prior data from up to eighteen years before the COVID crash. We also observed the distinct LPPLS bubble patterns in other same six subfigures in Figure 3 as in Figure 3 (a), including (b) DJIA, (c) NASDAQ, (e) DAX, (g) CSI300, (i) BSESN, and (j) BOVESPA. Once again, the same three remaining subfigures: (d) FTSE, (f) NIKKEI, and (h) HSI, do not show any detectable signature of the LPPLS model during the COVID crash, indicating that there is no endogenous positive bubble present in these three price trajectories. Furthermore, from Figure 3, we can not observe any LPPLS negative bubble pattern in the price trajectories of the 10 indexes during the COVID crash, which represents that these price trajectories did not present the two significant features of negative endogenous bubbles, that is, the super-exponential decline and the accelerating log-periodic oscillations, and the regime changes from the decline trend to growth trend were exogenous resulting from the external shocking events. The above observation results are consistent with the findings in Figure 2.

It can be also observed that the positive and negative bubble status measured by the confidence indicator in Figure 3 based on the weekly data may be significantly different from the results shown in Figure 2 based on the daily data. For example, the positive bubble clusters of DJIA index in Figure 3 (b) have a significantly different shape compared with the results in Figure 2 (b). The confidence indicator of positive bubble has a peak value of 21.6% in Figure 2 (b), indicating that the 27 out of 125 fitting windows can successfully pass the filter conditions, while this value in Figure 3 (b) is just 3.2%, that is, only the 4 out of 125 fitting have the signature of LPPLS model. This is a reasonable phenomenon because an asset price trajectory may exhibit a super-exponential growth in a short-term time scale, while it may present an exponential growth in a long-term scale, and vice versa.

Table 2 summarizes the statistics of positive bubble detection results for the 10 stock market indexes based on weekly data during 2020 global stock market crash from February to April 2020. Since the weekly price trajectory is extracted from the daily trading price trajectory, the dates and prices of peaks and valleys during the 2020 stock market crash may be slightly different from the corresponding values in Table 1.

Compared with a time series studied on a shorter time interval for a given trajectory, time series on a longer-time interval has smoother shape and fewer variations, so the LPPLS confidence indicator is more robust on the long-interval time series than that from the short-interval time series. Therefore this study has adopted the less conservative 2% confidence indicator value as the threshold for the weekly data to classify the type of crashes. From Table 2, the crashes in the SP500, DJIA, NASDAQ, DAX, CSI300, BSESN, and BOVESPA are seen as endogenous, while the crashes in the FTSE, NIKKEI, and HSI are exogenous. The classification results based on the weekly data in Table 2 are consistent with those based on the daily data in Table 1.

Table 2: Statistics of positive bubble detection based on weekly data during the 2020 global stock market crash.

| Index | Peak Date | Peak Price | Valley date | Valley Price | Crash Size | Peak CI | Type of Crash* |
|---|---|---|---|---|---|---|---|
| SP500 | 2/19/2020 | 3386.1 | 3/18/2020 | 2398.1 | 29.2% | 16.0% | Endogenous |
| DJIA | 2/19/2020 | 29348.0 | 3/18/2020 | 19898.92 | 32.2% | 3.2% | Endogenous |
| NASDAQ | 2/19/2020 | 9817.2 | 3/18/2020 | 6989.84 | 28.8% | 14.4% | Endogenous |
| FTSE | 2/11/2020 | 7499.4 | 3/17/2020 | 5294.9 | 29.4% | 0.0% | Exogenous |
| DAX | 2/19/2020 | 13789.0 | 3/18/2020 | 8441.71 | 38.8% | 3.2% | Endogenous |
| NIKKEI | 2/14/2020 | 23687.6 | 3/16/2020 | 17002.04 | 28.2% | 0.0% | Exogenous |
| CSI300 | 3/6/2020 | 4138.5 | 3/20/2020 | 3653.22 | 11.7% | 2.4% | Endogenous |
| HSI | 2/18/2020 | 27530.2 | 3/24/2020 | 22663.49 | 17.7% | 0.0% | Exogenous |
| BSESN | 2/12/2020 | 41565.9 | 3/27/2020 | 29815.59 | 28.3% | 10.4% | Endogenous |
| BOVESPA | 2/18/2020 | 114977.0 | 3/20/2020 | 67069 | 41.7% | 12.0% | Endogenous |

*Note: To ensure the classification robustness, the 2% confidence indicator value is used here as the threshold.

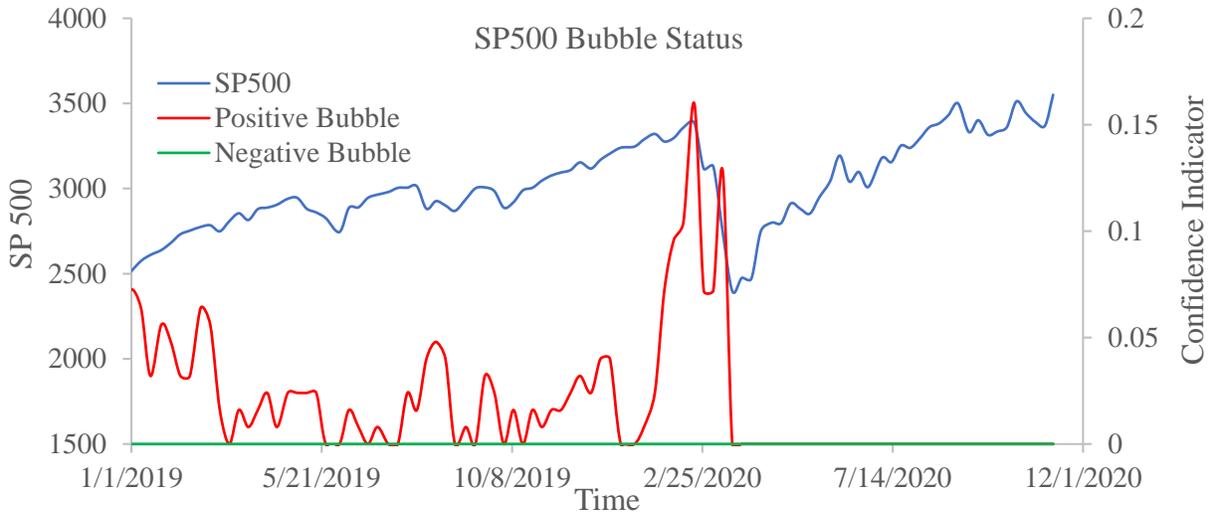

(a)

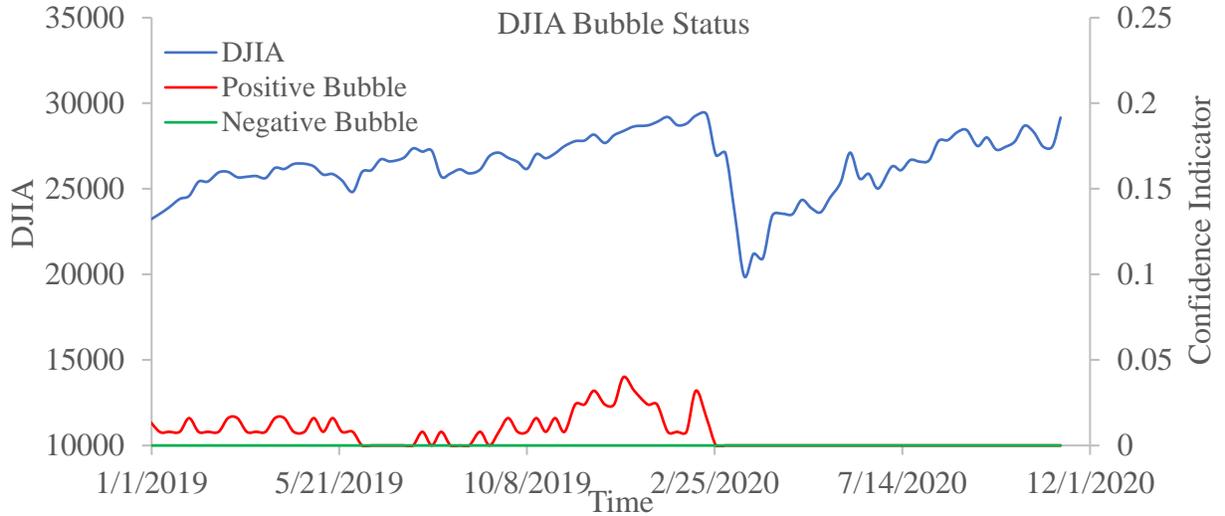

(b)

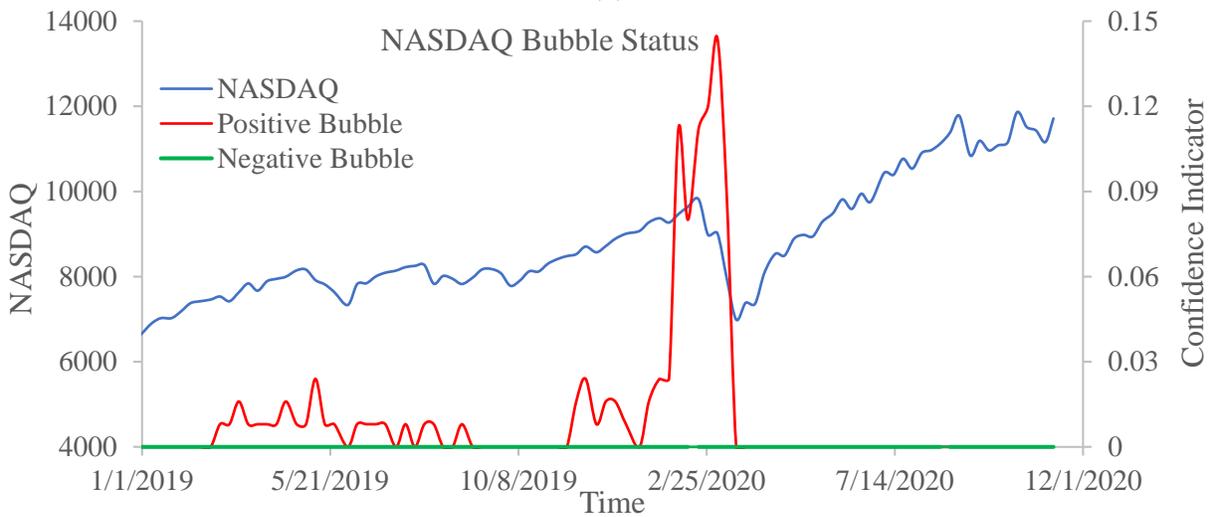

(c)

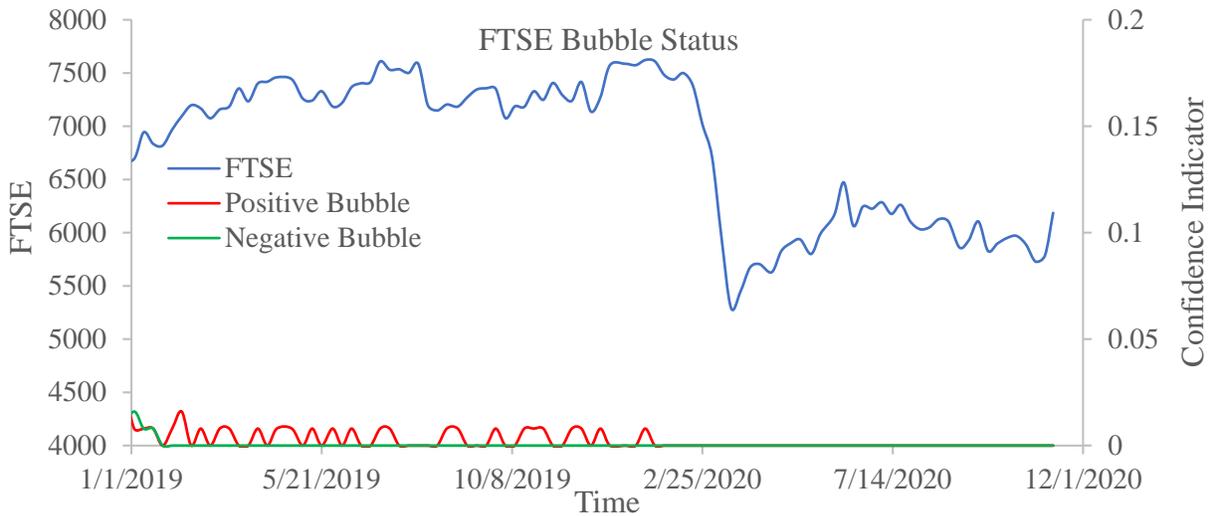

(d)

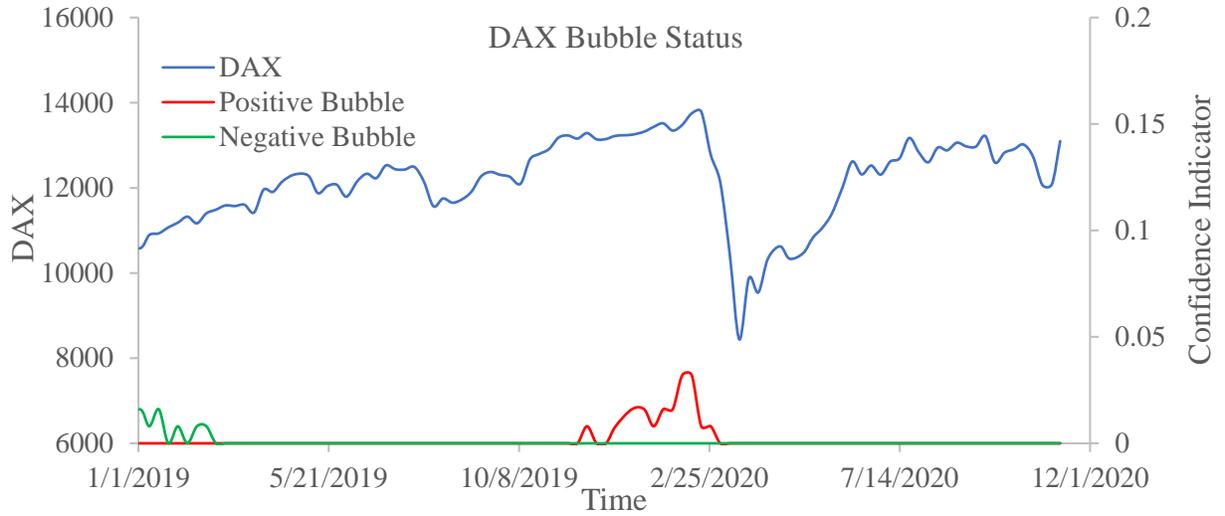

(e)

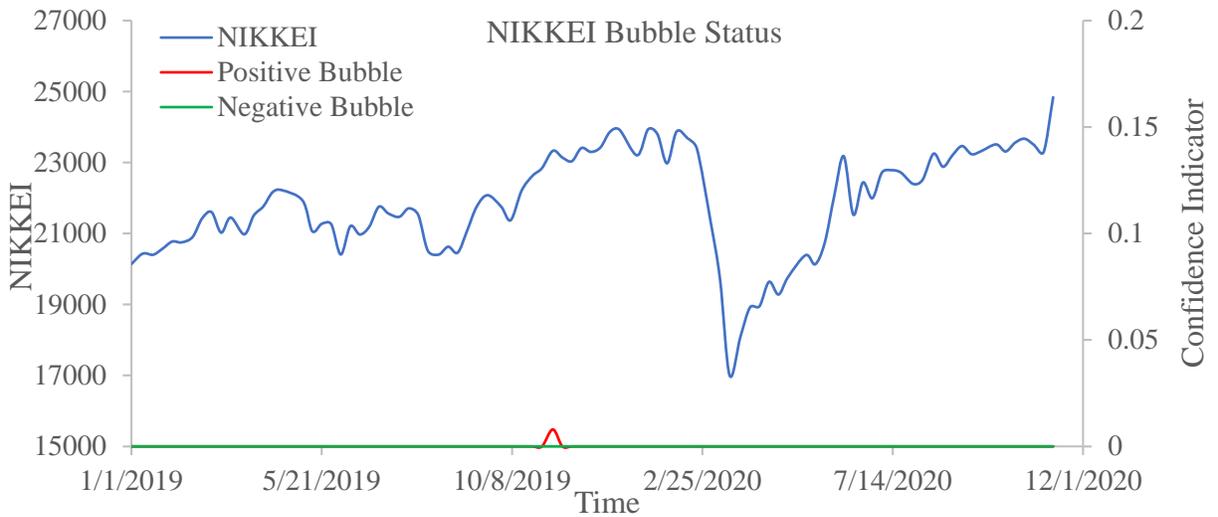

(f)

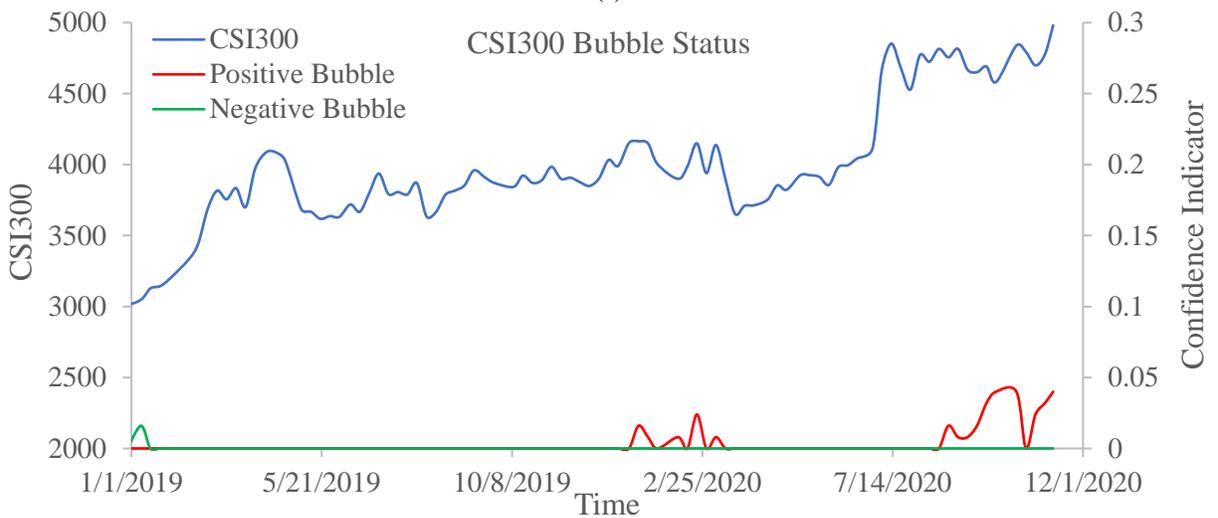

(g)

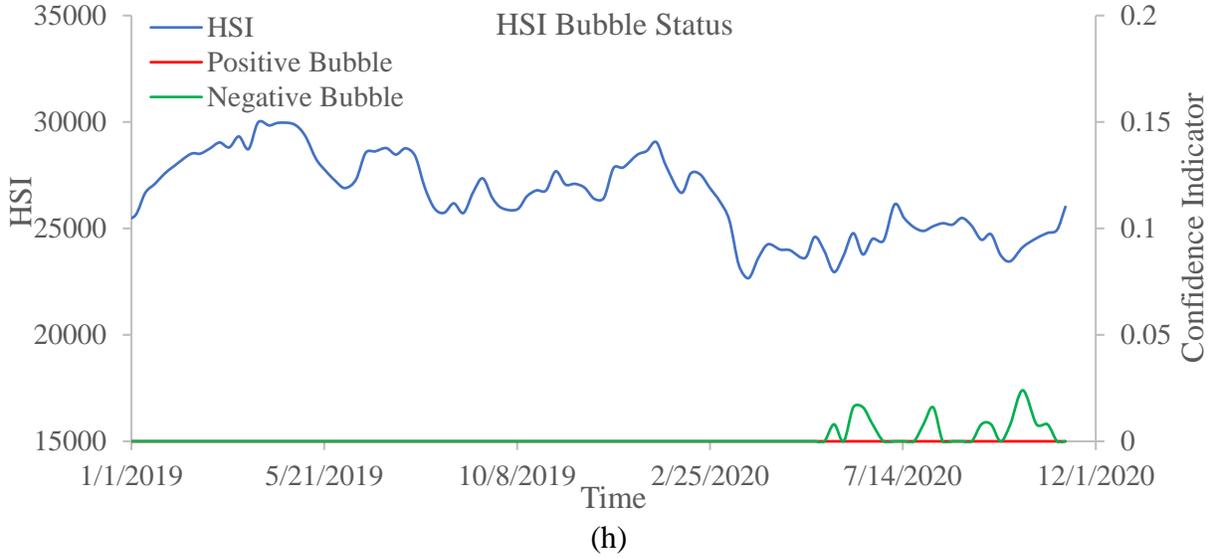

(h)

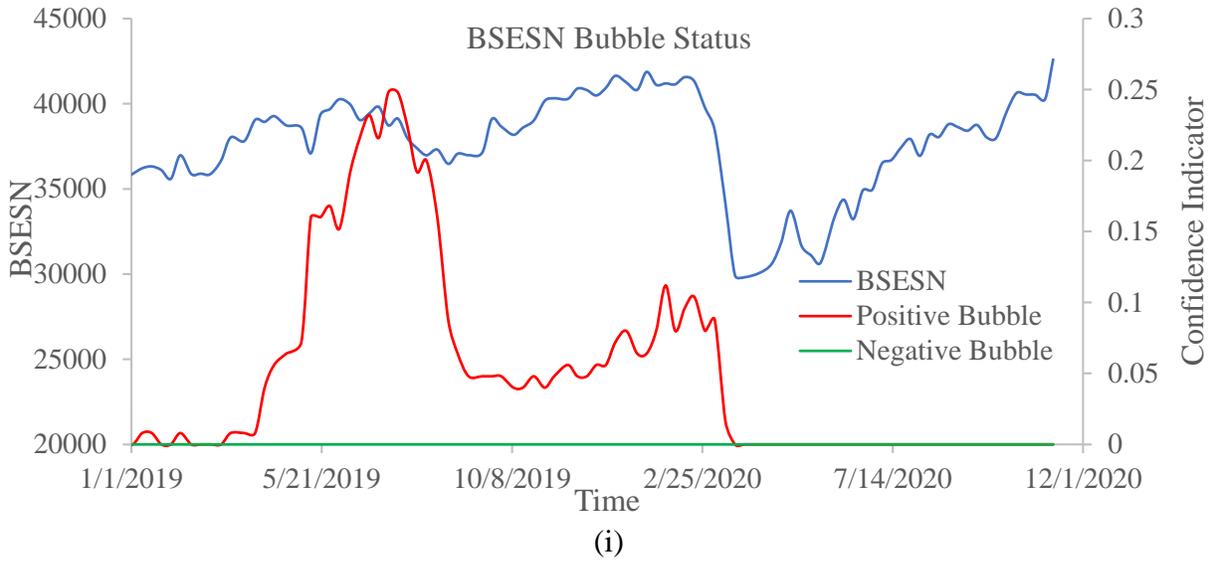

(i)

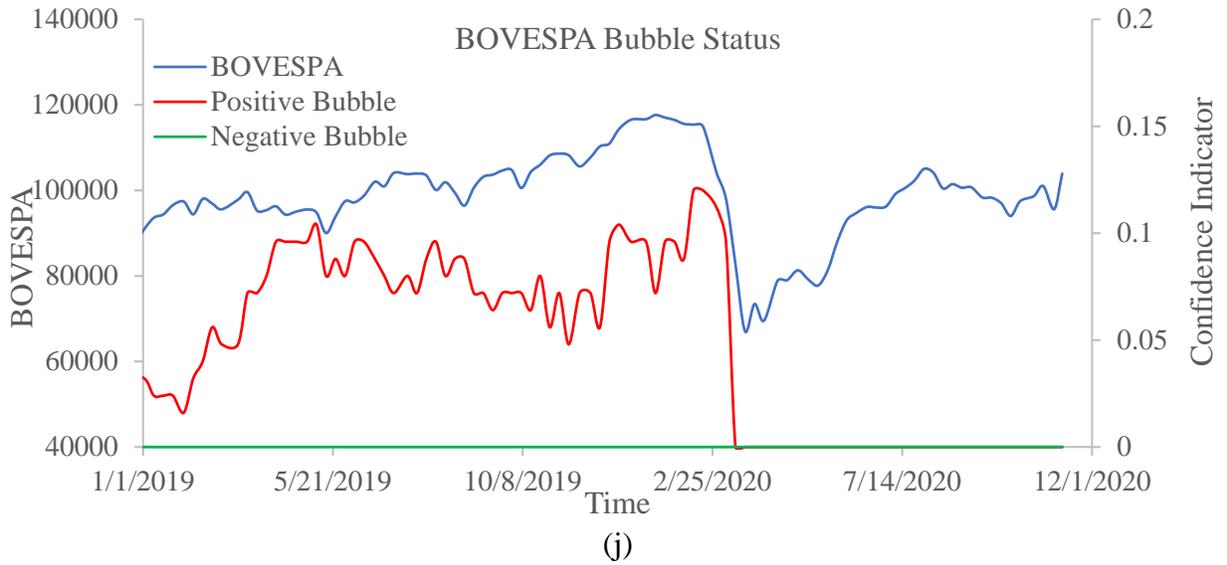

(j)

Figure 3. LPPLS confidence indicator for positive bubbles is shown in red and negative bubble in green (right scale) along with the index price in blue (left scale) for the 10 major stock market indexes based on weekly data from January 2019 to November 2020.

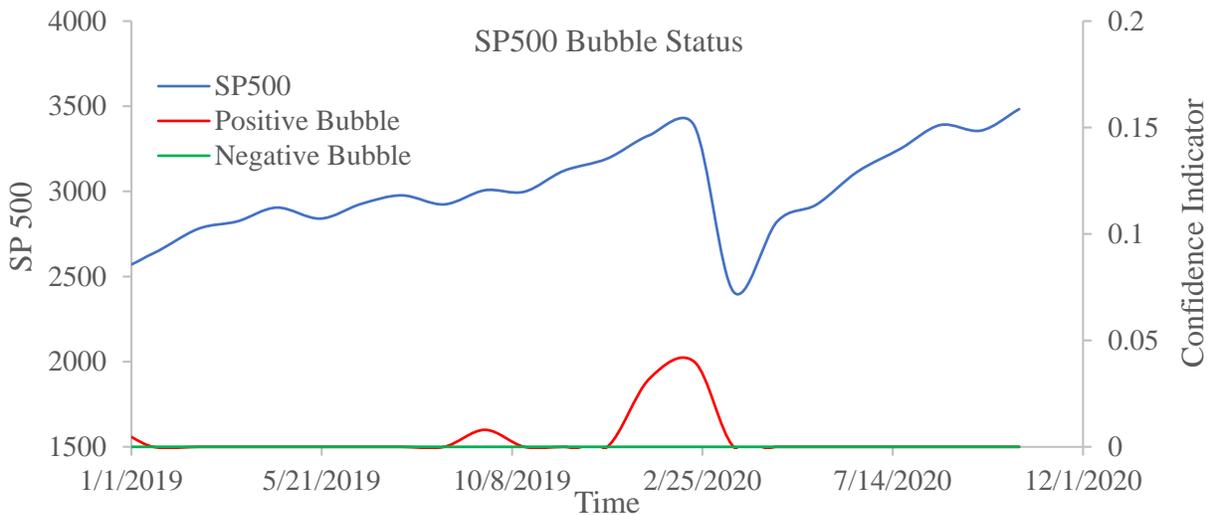

Figure 4. LPPLS confidence indicator for positive bubbles is shown in red and negative bubble in green (right scale) along with the index price in blue (left scale) for the SP500 stock market index based on the monthly data from January 2019 to November 2020.

### 3.3 Bubble Identification Based on Monthly Data

We adopted the monthly data in this section to investigate the 2020 global stock market crash. To generate the monthly time series of price trajectory, we extracted the data in the interval of the 21 trading days from the daily trading time series. To calculate the monthly confidence indicator, the length of time windows $t_2 - t_1$ is shrunk from 650 data points to 30 data points in steps of 5 data points based on the monthly time series, and the endpoint $t_2$ is moved from October 15, 2020 through January 17, 2019. The monthly confidence indicator can detect the bubble status for a specified time point by using the previous 30 to 650 data points, which corresponds to about three and a half years to seventy-seven years. Due to the data limitation, most of stock markets did not exist about half a century ago; we only studied the bubble status of the S&P 500 index.

Figure 4 presents the LPPLS confidence indicator for positive bubbles in red and negative bubble in green along with the index price in blue for the S&P5 00 index based on monthly data from January 17, 2019 to October 15, 2020. From Figure 4, we can see an obvious cluster of positive bubbles between January 17, 2020 and February 19, 2020 with the peak value of 4% on February 19, 2020. This shows that, for the S&P 500 index, the LPPLS bubble patterns observed in the 2020 global stock market crash on daily and weekly scales have persisted on the monthly scale -- this positive endogenous bubble can be detected even using prior data from up to seventy-seven years before the 2020 stock market crash. Not surprisingly, there is no LPPLS negative bubble pattern presented in the price trajectory of the S&P 500 index during the COVID crash.

## 4, Conclusions

In this study, we applied the LPPLS methodology using a multi-scale time series analysis paradigm to disclose the underlying mechanisms of the 2020 global stock market crash by analyzing the trajectories of 10 major stock market indexes from both developed and emergent stock markets, including the S&P 500, the DJIA, and the NASDAQ from the United State, the FTSE from the United Kingdom, the DAX from Germany, the NIKKEI from Japan, the CSI 300 from China, the HSI from Hong Kong, the BSESN from India, and the BOVESPA from Brazil. In order to effectively distinguish between endogenous crash and exogenous crash in stock market, we have proposed using the LPPLS confidence indicator as a classification proxy because the LPPLS model, rooted on the transient super-exponential growth of asset trajectories from self-reinforcing cooperative herding and imitative behaviors through interactions between market participants involving long-term memory processes of an endogenous organization, can only detect the endogenous crashes.

We investigated the 2020 global stock market crash, also referred to as the COVID crash, from different time scales by adopting both the daily data and the weekly data, which can provide insights into the bubble status from approximate forty days to two and a half years, and from about ten months to eighteen years, respectively. Our results show that the distinct LPPLS bubble patterns of the super-exponential increase corrected by the accelerating logarithm-periodic oscillations have presented clearly in the price trajectories of seven indexes: S&P500, DJIA, NASDAQ, DAX, CSI300, BSESN, and BOVESPA -- signifying that large positive bubbles have formed endogenously prior to the 'COVID' crash and the subsequent crashes for these seven indexes are endogenous, stemming from the increasingly systemic instability of the stock markets, while the well-known external shocks, such as the COVID-19 pandemic, the corporate debt bubble, and the 2020 Russia–Saudi Arabia oil price war, only acted as sparks during the 2020 global stock market crash. In contrast, the obvious signatures of LPPLS model have not been observed in the price trajectories of the remaining three indexes: FTSE, NIKKEI, and HSI, indicating that the crashes in these three indexes are exogenous caused by external shocks such as the COVID-19 pandemic etc. – and hence perhaps only crashes in these three markets can be truly referred to as the COVID crash.

Besides the positive bubbles, we have also tried to detect negative bubbles among the 10 stock market indexes – but to no avail. This indicates that the regime changes from a bear market to a bull market in late March 2020 are exogenous, stemming from external factors. The unprecedented market and economy rescue efforts from federal reserves and central banks across the world in unison may have played a critical role in quelling the 2020 global stock market crash in the nick of time.

Lastly, we have used the monthly data to analyze the bubble status of the S&P 500 index based on the prior data from about three and a half years to seventy-seven years ago. The results indicate that the LPPLS positive bubble patterns persisted in the S&P 500 index during the 2020 stock market crash -- showing that the positive endogenous bubble can be detected even using prior data from up to seventy-seven years before the 'COVID' crash.

In this work, we propose a novel classification method to unravel the underlying mechanisms of a stock crash – be it endogenous or exogenous, by adapting the log-periodic power law singularity (LPPLS) methodology. As an example, we have dissected the 2020 global stock market crash using a multi-resolution analysis paradigm, and successfully sifted out seven less COVID related endogenous crashes, and three more COVID related exogenous crashes. This new classification method can also be used to analyze crash types and regime changes of price trajectories in other financial markets and/or assets.

## Acknowledgment

The work is supported by the Faculty Research Initiative Grant as well as the New Faculty Start-Up Funds at the University of Wisconsin-Stout. The authors would like to thank the Blugold Supercomputing Cluster (BGSC) at the University of Wisconsin-Eau Claire.

## References

Albuquerque, R., Koskinen, Y., Yang, S., & Zhang, C. (2020). Resiliency of environmental and social stocks: an analysis of the exogenous COVID-19 market crash. *The Review of Corporate Finance Studies, 9*(3), 593-621.

Blanchard, O. J., & Watson, M. W. (1982). *Bubbles, rational expectations and financial markets.* Paper presented at the Crises in the Economic and Financial Structure, Lexington, MA.

Bothmer, H.-C., & Meister, C. (2003). Predicting critical crashes? A new restriction for the free variables. *Physica A: Statistical Mechanics and its Applications, 320*, 539-547.

Coy, P. (2020). The Great Coronavirus Crash of 2020 Is Different. Retrieved from https://www.bloomberg.com/news/articles/2020-03-19/the-great-coronavirus-crash-of-2020-is-different

Demirer, R., Demos, G., Gupta, R., & Sornette, D. (2019). On the predictability of stock market bubbles: evidence from LPPLS confidence multi-scale indicators. *Quantitative Finance, 19*(5), 843-858.

Demos, G., & Sornette, D. (2017). Birth or burst of financial bubbles: which one is easier to diagnose? *Quantitative Finance, 17*(5), 657-675.

Drozdz, S., Ruf, F., Speth, J., & Wójcik, M. (1999). Imprints of log-periodic self-similarity in the stock market. *The European Physical Journal B-Condensed Matter and Complex Systems, 10*(3), 589-593.

FED. (2020). Coronavirus Disease 2019 (COVID-19). Retrieved from https://www.federalreserve.gov/covid-19.htm

Feigenbaum, J. A., & Freund, P. G. (1996). Discrete scale invariance in stock markets before crashes. *International Journal of Modern Physics B, 10*(27), 3737-3745.

Filimonov, V., Demos, G., & Sornette, D. (2017). Modified profile likelihood inference and interval forecast of the burst of financial bubbles. *Quantitative Finance, 17*(8), 1167-1186.

Filimonov, V., & Sornette, D. (2013). A stable and robust calibration scheme of the log-periodic power law model. *Physica A: Statistical Mechanics and its Applications, 392*(17), 3698-3707. doi:10.1016/j.physa.2013.04.012

FRED. (2020). 10-Year Treasury Constant Maturity Rate. Retrieved from https://fred.stlouisfed.org/series/DGS10

Gerlach, J.-C., Demos, G., & Sornette, D. (2019). Dissection of Bitcoin's Multiscale Bubble History from January 2012 to February 2018. *Royal Society Open Science, 6*, 180643.

Guardian. (2020). Global stock markets post biggest falls since 2008 financial crisis. Retrieved from https://www.theguardian.com/business/2020/mar/09/global-stock-markets-post-biggest-falls-since-2008-financial-crisis

Hansen, N., Ostermeier, A., & Gawelczyk, A. (1995). *On the Adaptation of Arbitrary Normal Mutation Distributions in Evolution Strategies: The Generating Set Adaptation.* Paper presented at the the Sixth International Conference on Genetic Algorithms, San Francisco, CA.

Huang, Y., Johansen, A., Lee, M., Saleur, H., & Sornette, D. (2000). Artifactual log-periodicity in finite size data: Relevance for earthquake aftershocks. *Journal of Geophysical Research: Solid Earth, 105*(B11), 25451-25471.

Ide, K., & Sornette, D. (2002). Oscillatory finite-time singularities in finance, population and rupture. *Physica A: Statistical Mechanics and its Applications, 307*(1-2), 63-106.

Ihrig, J., Weinbach, G. C., & Wolla, S. A. (2020). COVID-19's Effects on the Economy and the Fed's Response. *Page One Economics®*.

Japantimes. (2020). G20 finance chiefs and central bankers to develop COVID-19 action plan. Retrieved from https://www.japantimes.co.jp/news/2020/03/25/business/economy-business/g20-covid-19-action-plan/#.XoQ8bi-ZNp9

Jiang, Z.-Q., Zhou, W.-X., Sornette, D., Woodard, R., Bastiaensen, K., & Cauwels, P. (2010). Bubble diagnosis and prediction of the 2005–2007 and 2008–2009 Chinese stock market bubbles. *Journal of Economic Behavior & Organization, 74*(3), 149-162.

Johansen, A., Ledoit, O., & Sornette, D. (2000). Crashes as critical points. *International Journal of Theoretical and Applied Finance, 3*(02), 219-255.

Johansen, A., & Sornette, D. (2010). Shocks, crashes and bubbles in financial markets. *Brussels Economic Review, 53*(2), 201-253.

Johansen, A., Sornette, D., & Ledoit, O. (1999). Predicting financial crashes using discrete scale invariance. *Journal of Risk, 1*(4), 5-32.

Li, C. (2017). Log-periodic view on critical dates of the Chinese stock market bubbles. *Physica A: Statistical Mechanics and its Applications, 465*, 305-311.

Lin, L., Ren, R. E., & Sornette, D. (2014). The volatility-confined LPPL model: A consistent model of 'explosive' financial bubbles with mean-reverting residuals. *International Review of Financial Analysis, 33*, 210-225. doi:10.1016/j.irfa.2014.02.012

Lynch, D. J. (2020). Fears of corporate debt bomb grow as coronavirus outbreak worsens. Retrieved from https://www.washingtonpost.com/business/2020/03/10/coronavirus-markets-economy-corporate-debt/

Mogi, C. (2020). Central Banks at Full Throttle Buying Bonds to Tame Markets. Retrieved from https://web.archive.org/web/20200330011208if_/https://finance.yahoo.com/news/boj-rba-spend-10-billion-023232886.html

RBI. (2020). RBI to conduct variable rate Term Repos of ₹1,00,000 crores. Retrieved from https://www.rbi.org.in/Scripts/BS_PressReleaseDisplay.aspx

Shu, M. (2019). *Identification and Forecasts of Bubbles and Crashes in Stock Market.* (Ph.D.), State University of New York at Stony Brook, ProQuest Dissertations Publishing. (13877774)

Shu, M., & Zhu, W. (2019). Diagnosis and Prediction of the 2015 Chinese Stock Market Bubble. *arXiv preprint arXiv:1905.09633*.

Shu, M., & Zhu, W. (2020a). Detection of Chinese stock market bubbles with LPPLS confidence indicator. *Physica A: Statistical Mechanics and its Applications, 557*, 124892.

Shu, M., & Zhu, W. (2020b). Real-time prediction of Bitcoin bubble crashes. *Physica A: Statistical Mechanics and its Applications, 548*, 124477.

Sornette, D. (1998). Discrete-scale invariance and complex dimensions. *Physics Reports, 297*(5), 239-270.

Sornette, D. (2003). Critical market crashes. *Physics Reports, 378*(1), 1-98. doi:10.1016/s0370-1573(02)00634-8

Sornette, D., Demos, G., Zhang, Q., Cauwels, P., Filimonov, V., & Zhang, Q. (2015). Real-time prediction and post-mortem analysis of the Shanghai 2015 stock market bubble and crash. *Journal of Investment Strategies, 4*(4), 77–95.

Sornette, D., & Johansen, A. (1997). Large financial crashes. *Physica A: Statistical Mechanics and its Applications, 245*(3), 411-422.

Sornette, D., & Johansen, A. (2001). Significance of log-periodic precursors to financial crashes. *Quantitative Finance, 1*(4), 452-471.

Sornette, D., Johansen, A., & Bouchaud, J.-P. (1996). Stock market crashes, precursors and replicas. *Journal de Physique I, 6*(1), 167-175.

Sornette, D., Woodard, R., & Zhou, W.-X. (2009). The 2006–2008 oil bubble: Evidence of speculation, and prediction. *Physica A: Statistical Mechanics and its Applications, 388*(8), 1571-1576.

Sornette, D., & Zhou, W. X. (2002). The US 2000-2002 market descent: How much longer and deeper? *Quantitative Finance, 2*(6), 468-481.

WHO. (2020). Timeline: WHO's COVID-19 response. Retrieved from https://www.who.int/emergencies/diseases/novel-coronavirus-2019/interactive-timeline#!

Yan, W., Woodard, R., & Sornette, D. (2010). Diagnosis and prediction of tipping points in financial markets: Crashes and rebounds. *Physics Procedia, 3*(5), 1641-1657.

Zhou, W., & Sornette, D. (2003). 2000–2003 real estate bubble in the UK but not in the USA. *Physica A: Statistical Mechanics and its Applications, 329*(1), 249-263.

Zhou, W., & Sornette, D. (2006). Is there a real-estate bubble in the US? *Physica A: Statistical Mechanics and its Applications, 361*(1), 297-308.

Zhou, W., & Sornette, D. (2008). Analysis of the real estate market in Las Vegas: Bubble, seasonal patterns, and prediction of the CSW indices. *Physica A: Statistical Mechanics and its Applications, 387*(1), 243-260.